\documentclass[aps,reprint,notitlepage,floatfix,superscriptaddress,prl]{revtex4-1}

\usepackage[pdftex]{graphicx}
\usepackage{amsmath,amssymb,amstext}
\usepackage{mathrsfs}
\usepackage{multirow}
\usepackage{comment}
\usepackage{bbold}
\usepackage{dsfont}
\usepackage{color}
\usepackage{hyperref}
\usepackage{tikz}
\usepackage{braket}
\usepackage[colorinlistoftodos]{todonotes}
\usepackage{soul}
\renewcommand{\eqref}[1]{Eq.~(\ref{#1})}
\newcommand{\figref}[1]{Fig.~\ref{#1}}

\begin{document}
\title{Ultrafast Measurement of Energy-Time Entanglement with an Optical Kerr Shutter}

\author{Andrew R. Cameron}
\email{ar3camer@uwaterloo.ca}
\affiliation{Institute for Quantum Computing, University of Waterloo, Waterloo, ON N2L 3G1, Canada}
\affiliation{Department of Physics \& Astronomy, University of Waterloo,Waterloo, ON N2L 3G1, Canada}

\author{Kate L. Fenwick}
\email{kfenw032@uottawa.ca}
\affiliation{National Research Council of Canada, 100 Sussex Drive, Ottawa, ON K1A 0R6, Canada}
\affiliation{Department of Physics, University of Ottawa, Advanced Research Complex, 25 Templeton Street, Ottawa, ON K1N 6N5, Canada }

\author{Sandra W. L. Cheng}
\affiliation{Institute for Quantum Computing, University of Waterloo, Waterloo, ON N2L 3G1, Canada}
\affiliation{Department of Physics \& Astronomy, University of Waterloo,Waterloo, ON N2L 3G1, Canada}

\author{Sacha Schwarz}
\affiliation{Institute for Quantum Computing, University of Waterloo, Waterloo, ON N2L 3G1, Canada}
\affiliation{Department of Physics \& Astronomy, University of Waterloo,Waterloo, ON N2L 3G1, Canada}
\affiliation{Infinite Potential Laboratories LP, 485 Wes Graham Way, Waterloo, ON N2L A07, Canada}

\author{Benjamin MacLellan}
%\email{benjamin.maclellan@uwaterloo.ca}
\affiliation{Institute for Quantum Computing, University of Waterloo, Waterloo, ON N2L 3G1, Canada}
\affiliation{Department of Physics \& Astronomy, University of Waterloo,Waterloo, ON N2L 3G1, Canada}

\author{Philip J. Bustard}
\affiliation{National Research Council of Canada, 100 Sussex Drive, Ottawa, ON K1A 0R6, Canada}

\author{Duncan England}
\affiliation{National Research Council of Canada, 100 Sussex Drive, Ottawa, ON K1A 0R6, Canada}

\author{Benjamin Sussman}
\affiliation{National Research Council of Canada, 100 Sussex Drive, Ottawa, ON K1A 0R6, Canada}
\affiliation{Department of Physics, University of Ottawa, Advanced Research Complex, 25 Templeton Street, Ottawa, ON K1N 6N5, Canada }

\author{Kevin J. Resch}
\affiliation{Institute for Quantum Computing, University of Waterloo, Waterloo, ON N2L 3G1, Canada}
\affiliation{Department of Physics \& Astronomy, University of Waterloo,Waterloo, ON N2L 3G1, Canada}

\begin{abstract}
Recent experimental progress in quantum optics has enabled measurement of single photons on ultrafast timescales, beyond the resolution limit of single photon detectors. The energy-time degree of freedom has emerged as a promising avenue for quantum technologies, as entanglement between the frequency and temporal properties of two photons can be fully explored and utilized. Here, we implement optical Kerr shutters in single mode fibers to map out the sub-picosecond correlations of energy-time entangled photon pairs. These measurements, in addition to joint spectral measurements of the photon pair state, are used to verify entanglement by means of the violation of a time-bandwidth inequality.
\end{abstract}

\maketitle

The energy-time degree of freedom is important for many quantum technologies, including quantum networks \cite{quantum_network_citation,tannous2023fully}, optical quantum computers \cite{xanadu}, and quantum sensing \cite{PhysRevA.101.053808}. This degree of freedom is useful due to its intrinsic robustness against decoherence for long-distance transmission of quantum information \cite{cuevas_long-distance_2013}, increasing imaging resolution via interferometric techniques \cite{brown2022interferometric}, and for realizing high-dimensional entangled quantum states \cite{steinlechner_distribution_2017}. Energy-time entangled pairs of ultrafast photons (femtosecond-picosecond duration) are challenging to control and measure with sufficient resolution. Measuring in the sub-picosecond regime is particularly important because there has yet to be a single photon detector with comparable resolution. The highest detector resolution to date has been demonstrated with superconducting nanowire single-photon detectors (SNSPD), which have seen timing resolution on the order of a few picoseconds to tens of picoseconds, depending on the photon's frequency~\cite{korzh_demonstration_2020,doi:10.1063/5.0128129}. 

Fast gating of optical signals is commonly performed electronically by micromechanical switches  \cite{micromechanical_switch_4micro_s,mem2} or electro-optic modulators \cite{EOM_singlephotons} on nanosecond timescales. Optical gating has been used to surpass these timing restrictions, with resolutions of 450\,ps in ring resonators \cite{optical_switching_silicon}, 10\,ps in nonlinear optical loop mirrors \cite{NOLM_2011_entanglement}, and sub-picosecond with sum-frequency generation (SFG) \cite{allgaier_fast_2017}. 
Optical gating is paramount in the detection and control of sub-picosecond energy-time-entangled photon pairs~\cite{energy-time-characterization}. SFG optical gating has been used for time-resolved detection of energy-time entangled photon pairs \cite{PhysRevLett.101.153602}, which can exhibit correlations in time on the order of a few femtoseconds. SFG temporal measurements have been taken alongside spectral measurements to completely reconstruct a two-photon joint spectral amplitude~\cite{quantumfrog}. 

A promising alternative method for ultrafast optical gating is to use an optical Kerr shutter (OKS)~\cite{england_perspectives_2021}. This method relies on the optical Kerr effect which can occur in any material, including those which are centrosymmetric. This makes it suitable for integration in standard single-mode fiber (SMF), where spatial overlap between the signal and pump is easy to achieve. In an OKS, the transient birefringence induced by a strong laser pulse will rotate the polarization of a photon pulse only where the two pulses temporally overlap in the Kerr medium. Picosecond and sub-picosecond Kerr gating has been shown for classical applications, such as optical communications~\cite{whitaker1991all, yamada1995subpicosecond, moller2003ultrahigh}. Previous demonstrations for quantum applications have shown near-unit efficiency operation of the Kerr shutter in YAG crystal~\cite{kupchak2017time} and short (10\,cm) pieces of SMF on picosecond timescales~\cite{terahertz,bouchard2021achieving,bouchard2022quantum,bouchard2023measuring}. In these cases, the group velocity walk-off between the signal and pump in the SMF was exploited to fully switch the polarization qubit of a single photon for quantum communication and information applications. Operation with a photon and pump pulse close in wavelength reduces walk-off and has been used to demonstrate Kerr shutter resolution as low as 305\,fs~\cite{fenwick2020carving}; however, this operation regime introduces noise from pump self-phase modulation and cannot easily be operated above 30\%. Walk-off reduction has also been implemented using photonic crystal fiber (PCF) as the $\chi^{(3)}$ medium to match the group velocities of both pulses~\cite{sagnac_kerr_switch}.

In this work, we implement an OKS for each entangled photon and demonstrate fast gated measurements with $320 \pm 30$~fs and $290 \pm 30$~fs resolution for the signal and idler photons, respectively. We characterize photon-pair correlations in time without the need for an interferometric setup. Unlike previous experiments using SFG for optical gating, no frequency conversion of the photons is required and the only difference between switched and unswitched photons is their polarization. Here, we will be using this polarization rotation to measure time correlations; however, polarizing optics after an OKS could easily be used to reroute photons with sub-picosecond resolution for use in other quantum information protocols.  

%%%%%%%%%%%%%%%%%%%%%%%%%%%%%%%%%%%%%%%%%%%%%%%%%%%%%%%%%%%%%%%%%%%%%%%%%%%%%%%%%%%%%%%%%%%%%%%%%%%%%%%%%%%%%%%%%%%%%%%%%%%%%%

Uncertainty relations can be used to detect energy-time entanglement through temporal and spectral measurements. Two separable photons labeled signal (s) and idler (i) must satisfy the inequality~\cite{insep_criteria,insep_criteria_2}

\begin{equation}
\Delta (\omega_s + \omega_i )\Delta (t_s - t_i) \geq 1 ,
    \label{time_bandwidth_inequality}
\end{equation}

\noindent
where $\omega$ corresponds to frequency, $t$ corresponds to time of arrival, and $\Delta (\omega_s + \omega_i )\: (\Delta (t_s - t_i))$ signifies standard deviation in the sum of their frequencies (difference of their detection times) \cite{shalm_three-photon_2013}. The quantities in Eq.~\ref{time_bandwidth_inequality} can be determined experimentally by measuring the joint spectral intensity (JSI) and joint temporal intensity (JTI) of a two-photon system. Violating this inequality is a sufficient condition for witnessing entanglement. 

%This inequality stems from the position-momentum uncertainty relation $\Delta x_i\Delta p_i \geq \hbar/2$ of particles with commutation relations $[x_i,p_j] = i\hbar \delta_{ij}$ for position and momentum operators $x$ and $p$. Photons travel at the speed of light, $c$, which provides a mapping of longitudinal position, $x$, to the time of arrival, $t = x/c$, and a mapping of longitudinal momentum, $p$, to the light frequency $\hbar \omega = cp$ \cite{shalm_three-photon_2013,energy-time-position-momentum}. 

\begin{figure}[ht!]
\centering
\includegraphics[trim={0 0 0 0},clip,width=0.9\columnwidth]{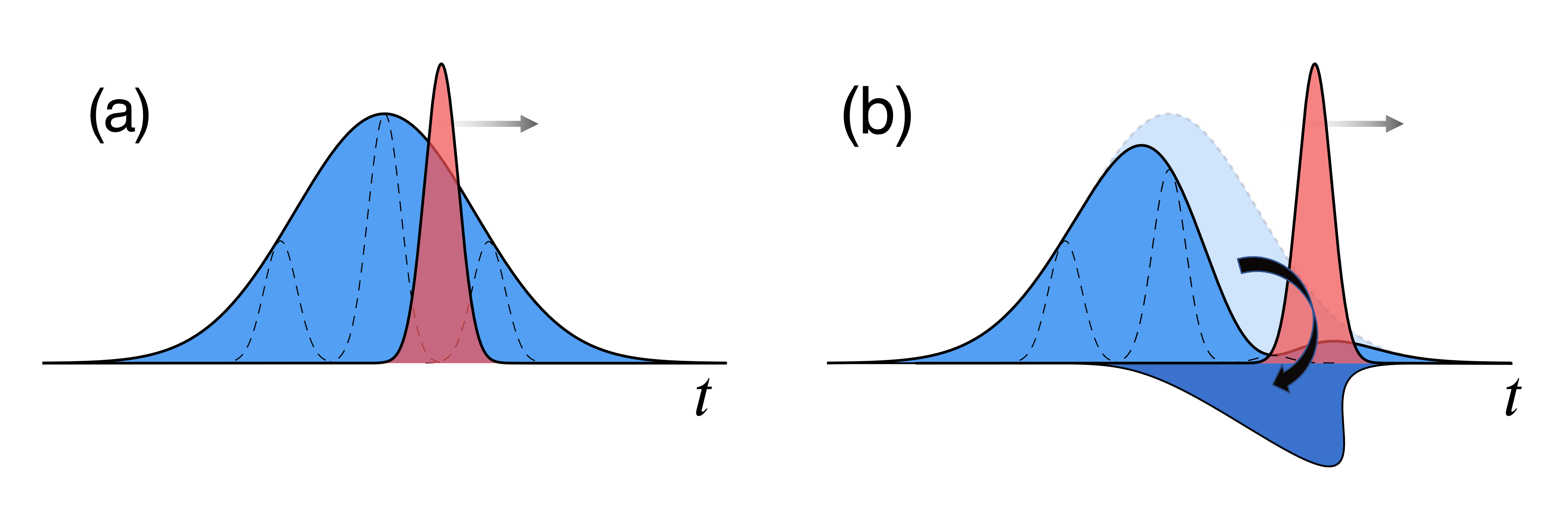}
\caption{Visualization of the main operating principle of an OKS. A 775\,nm pump pulse walk-off in a 35\,mm fiber in the reference frame of a 714\,nm photon (a) Initial temporal representations of the pump pulse relative to the photon. The broader photon distribution, shown in blue, is a combination of shorter photons created at varying times in the crystal, depicted with dotted lines. The pump pulse is shown in red. (b) Final temporal representation after the pump has swept through a section of the photon pulse, rotating its polarization. Delaying the pump pulse relative to the photon allows for gated measurement of different temporal segments of the photon.    
 }
\label{fig:walkoff}
\end{figure}

The overall two-photon temporal measurement resolution can be estimated by the quadrature sum of the pump pulse at each shutter, $\tau_{\text{p}}$, the walk-off between the pump and signal (idler) photon, $\Delta \tau_{s(i)}$, and the intrinsic width of the JTI about the $t_s = -t_i$ axis, $\Delta (t_s - t_i)_{\text{Int}}$, as

\begin{equation}
\Delta (t_s - t_i) = \sqrt{\Delta \tau^2_{s} + \Delta \tau^2_{i} + 2\tau^2_{\text{p}} + \Delta (t_s - t_i)_{\text{Int}}^2},
    \label{jti_width_model}
\end{equation}

\noindent
where $\Delta \tau_{s(i)} = L(v_{g_{s(i)}}^{-1} - v_{g_{\text{p}}}^{-1})$, $L$ is the length of the Kerr medium, $v_{g_{s(i)}}$ is the group velocity of the signal (idler) photon, and $v_{g_{\text{p}}}$ is the group velocity of the pump pulse. Note that the factor of two in front of the $\tau^2_p$ term accounts for the pump pulse width at each of the two shutters. Photons from spontaneous parametric downconversion (SPDC) have strong correlations in time of arrival and exhibit an intrinsic width $\Delta (t_s - t_i)_{\text{Int}}$. This width is modelled by first considering the spectral correlations and then using a double Fourier transform to the time domain. The walk-off terms in Eq.~\ref{jti_width_model}, arising from the difference in group velocity between the pump and photons, further increase the width of the JTI. The effect of walk-off on temporal resolution is visualized in \figref{fig:walkoff}. This figure also illustrates the photon's temporal distribution, consisting of short photon pulses with different possible arrival times.

%%%%%%%%%%%%%%%%%%%%%%%%%%%%%%%%%%%%%%%%%%%%%%%%%%%%%%%%%%%%%%%%%%%%%%%%%%%%%%%%%%%%%%%%%%%%%%%%%%%%%%%%%%%%%%%%%%%%%%%%%%%%%%

%\section{Experimental Method}

\begin{figure*}[ht!]
\centering
\includegraphics[trim={3cm 7cm 0 5cm},width=\textwidth]{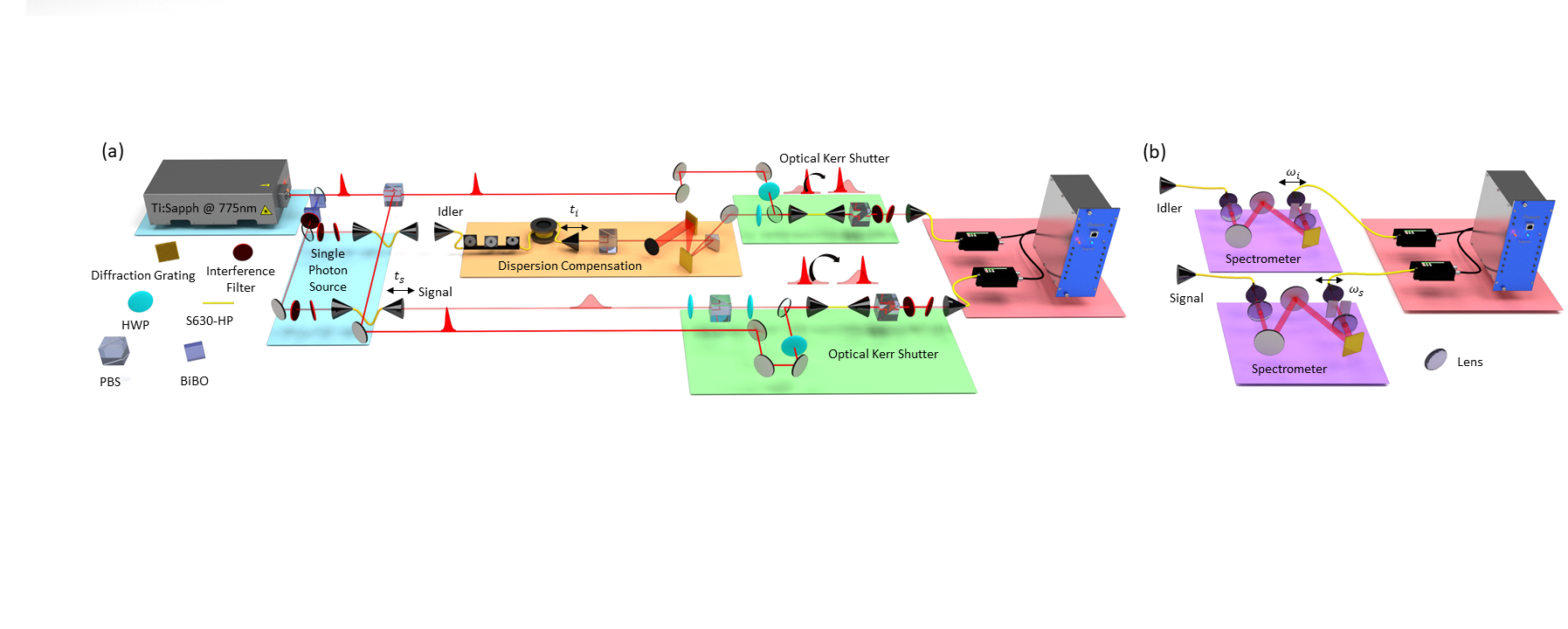}
\caption{Schematic of experimental setup. (a) A titanium-sapphire (Ti:Sapph) laser produces 775\,nm pulsed light with a repetition rate of 80\,MHz. The light is upconverted in a 2\,mm $\beta $-BiBO crystal to 387.5\,nm to pump the single photon source. Energy-time entangled photon pairs are generated by Type-I SPDC in a 5\,mm BiBO crystal. The 847\,nm idler photon is sent through a 21.2\,m fiber and a grating compressor for dispersion control of the idler photon and non-local dispersion compensation of the 714\,nm signal photon. Each photon is Kerr-gated by a strong pulse, picked-off from the output of the Ti:Sapph laser, within a 35\,mm piece of SMF (Thorlabs S630-HP). Coincidence detection of the output of each OKS enables measurement of the JTI of the two-photon entangled state. (b) Each photon from the source can alternatively be directed to a scanning monochromator for measurement of the JSI. In this configuration, a grating is used to spread the single photon's optical frequency components spatially, after which a single photon detector can be spatially scanned to detect a given frequency component.}
\label{fig:expSetup}
\end{figure*}

An illustration of the experiment is shown in \figref{fig:expSetup}. Optical pulses with 148\,fs full-width at half maximum (FWHM) from a titanium sapphire laser with an 80\,MHz repetition rate are guided to three parts of the experiment: an energy-time entangled photon source and two Kerr shutters. The pulsed light is first upconverted by a 2\,mm $\beta$-bismuth borate (BiBO) crystal to 387.5\,nm, spectrally filtered by a 0.1\,nm FWHM bandpass filter, and then downconverted in a 5\,mm BiBO crystal to produce energy-time entangled photon pairs with wavelengths 714\,nm and 847\,nm which we refer to as the signal and idler, respectively. Each photon is spectrally narrowed to a bandwidth of 6\,nm FWHM, after which we measure a signal count rate of $3.6\times 10^6 \text{s}^{-1}$, an idler count rate of $2.6\times 10^6 \text{s}^{-1}$, and a signal-idler coincidence count rate of  $4\times10^5\text{s}^{-1}$ using a 3\,ns coincidence window. Both photons are fiber-coupled and directed to (a) an OKS, or (b) a scanning monochromator. 

The two photons travel through different lengths of fiber to reach their respective OKS. Ideally, our final measurement of the temporal profile of each photon would reveal the true transform-limited widths in time; however, dispersion in each fiber stretches the temporal profile of each photon. The signal photon passes through 50\,cm of fiber which applies 10,910\,fs$^2$ of dispersion. The idler photon passes through 21.2\,m of fiber which applies 344,064\,fs$^2$ of dispersion. One way to compensate for the dispersion in each path would be to build a grating compressor after each fiber to apply negative dispersion to each photon. Optimal gratings with a near-Littrow configuration provide a total compressor transmission efficiency of around 65\% which drops further for shorter wavelength light \cite{lavoie_spectral_2013,grating_paper,walmsley_role_2001}, so to avoid unnecessary loss in the experiment we build a single grating compressor in the idler path. Due to the photon pair time-of-arrival correlations, the gratings can locally cancel the dispersion of the idler and non-locally cancel the dispersion of its partner signal photon \cite{dispersion_cancellation} by applying second order dispersion of $-(344,064 + 10910)\,\text{fs}^{2}$ on the idler photon. Experimentally, we adjust the grating position to minimize the temporal width of the JTI.

%The compressor applies 1740fs$^2$ of quadratic dispersion per mm of grating separation, and is fine tuned for dispersion cancellation by minimizing the width of the Kerr-gated photons in time.

Both the signal and idler photon polarizations are prepared along the horizontal axis, while the pump polarization is prepared at $45^\circ$ relative to the photons to maximize Kerr-rotation along the interaction region~\cite{terahertz}. Each photon is combined with a pump pulse on a dichroic mirror and directed into a 35\,mm SMF (Thorlabs S630-HP). Both devices have an equal input pump power of 800mW. Unlike bulk media nonlinear optical crystals, SMF has the advantage that two modes are confined to a small core on the order of 5\,$\mu$m in diameter which maintains high intensity to increase the nonlinear effect and also facilitates alignment. Pump and photon pulses were coupled into their respective SMFs, each with 40\% coupling efficiency. In each OKS, only the part of the photon pulse that overlaps with the pump pulse in the Kerr medium will have its polarization rotated. This portion of the photon pulse will transmit through the Glan-Taylor polarizing beamsplitter (PBS) at the output of each 35\,mm SMF. The pump pulse and photon are then separated by interference filters angle-tuned to pass the 6\,nm bandwidth of the gated photons. The pump is 61\,nm and 72\,nm away from the signal and idler photons in wavelength, respectively, which is an equal 32\,THz on either side of the pump; however, unwanted noise processes, such as self-phase modulation and Raman scattering, can generate pump noise in the spectral region of the single photons. As a result, tight spectral filtering is required for lowering background counts. 

%There is a trade off between efficiency, noise, and temporal resolution in an OKS. For high efficiency, we need a long fiber and high pump power. For low noise, we need a short fiber, low pump power, and large frequency difference between pump and photon pulses to limit self-phase modulation and Raman scattering producing noise photons with the same frequency as the output signal. Finally, for high time resolution, we need a short fiber and small frequency difference between pump and photon pulses. Clearly, we can't have all three parameters maximized simultaneously. In this work, we found a set of parameters, frequency difference $\Delta \omega = 32$\,THz, fiber length $L = 35$\,mm, and pump power $P = 800$\,mW, that present a compromise enabling high enough efficiency to observe photon-pair time correlations with sufficient resolution to detect energy-time entanglement. 

The light that transmits through the PBS and spectral filters, consisting of pump noise and Kerr-gated light from the single photons, is fiber-coupled and detected by avalanche photodiodes. Counts registered by each detector are time-tagged and analyzed with coincidence logic using a 3\,ns coincidence window. The relative time of arrival between each photon and its corresponding pump pulse is varied by stepper-motors on the output fiber couplers before each OKS and denoted $t_s$ and $t_i$ in \figref{fig:expSetup}(a). Sweeping both motors in a raster scan pattern and counting coincidences at each position fully maps out the JTI of the two-photon system. As shown in  \figref{fig:expSetup}(b), spectral measurements are made directly after the source with each photon fiber-coupled and routed directly to scanning monochromators, as in previous experiments \cite{energy-time-characterization,quantumfrog}. Similarly, raster scanning the motors through frequency measurements on each photon builds up the JSI. We note that it would be possible to work with the energy-time degree of freedom on slower timescales than the requirements of this experimental setup, but only at the cost of requiring higher frequency resolution in the scanning monochromators instead of high temporal resolution.

\begin{figure}[ht!]
\centering
\includegraphics[trim={1cm 0 6cm 0},clip=true,width=\columnwidth]{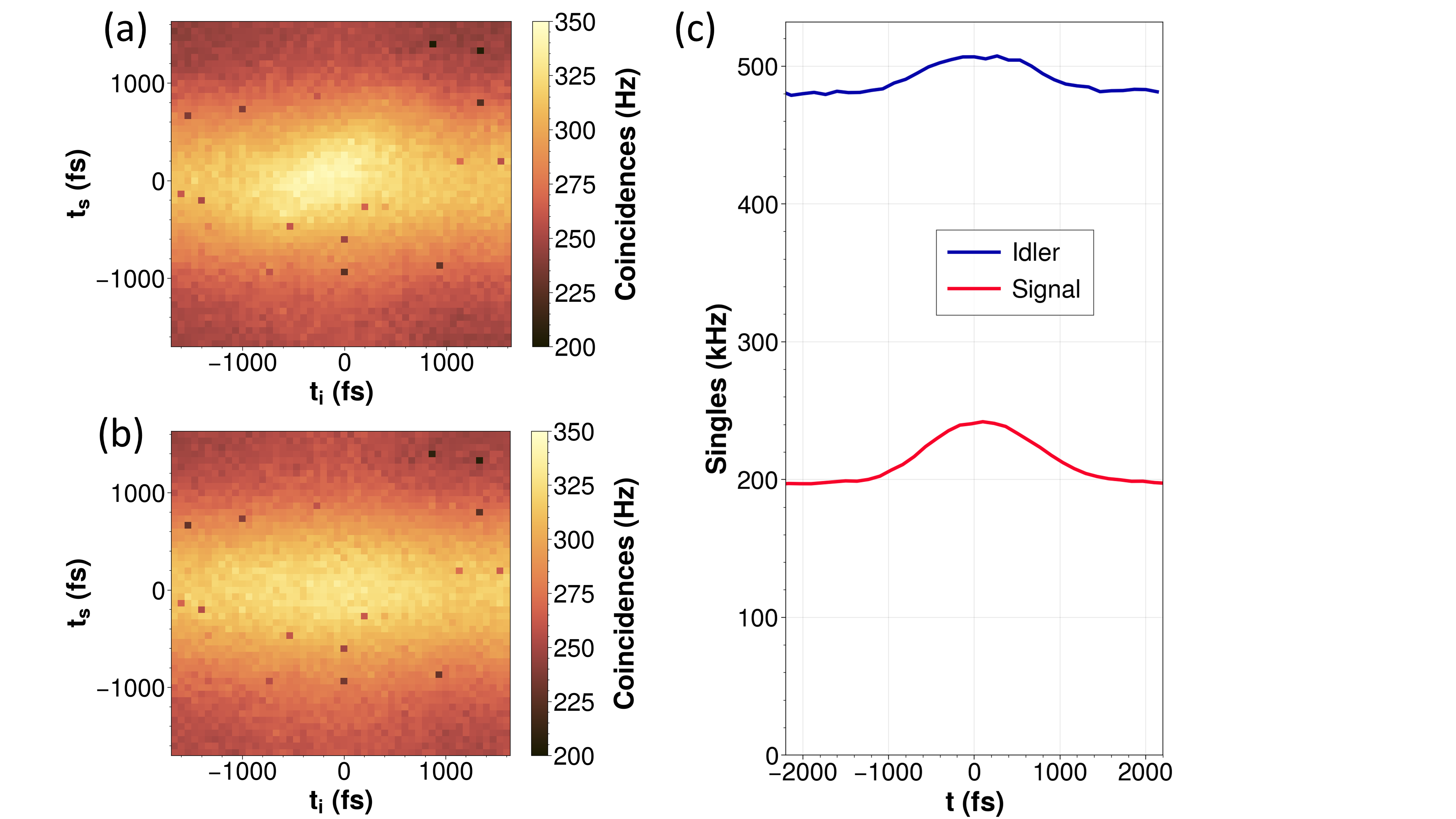}
\caption{Background subtraction in post processing. (a) Measured coincidences between two Kerr shutters at different relative gate delays $t_s$ and $t_i$. (b) Estimation of accidental coincidences and constant pump background, measured by adding 12.5ns of electronic delay between the two photon counting signals (corresponding to the repetition rate of the laser). The small number of artifacts that make a speckle pattern across the image are the result of a momentary fault in time tagging electronics. (c) Single counts at detectors after each OKS.
 }
\label{fig:background_subtraction}
\end{figure}

\begin{figure}[ht]
\centering
\includegraphics[trim={2cm 0 5cm 0},clip=true,width=\columnwidth]{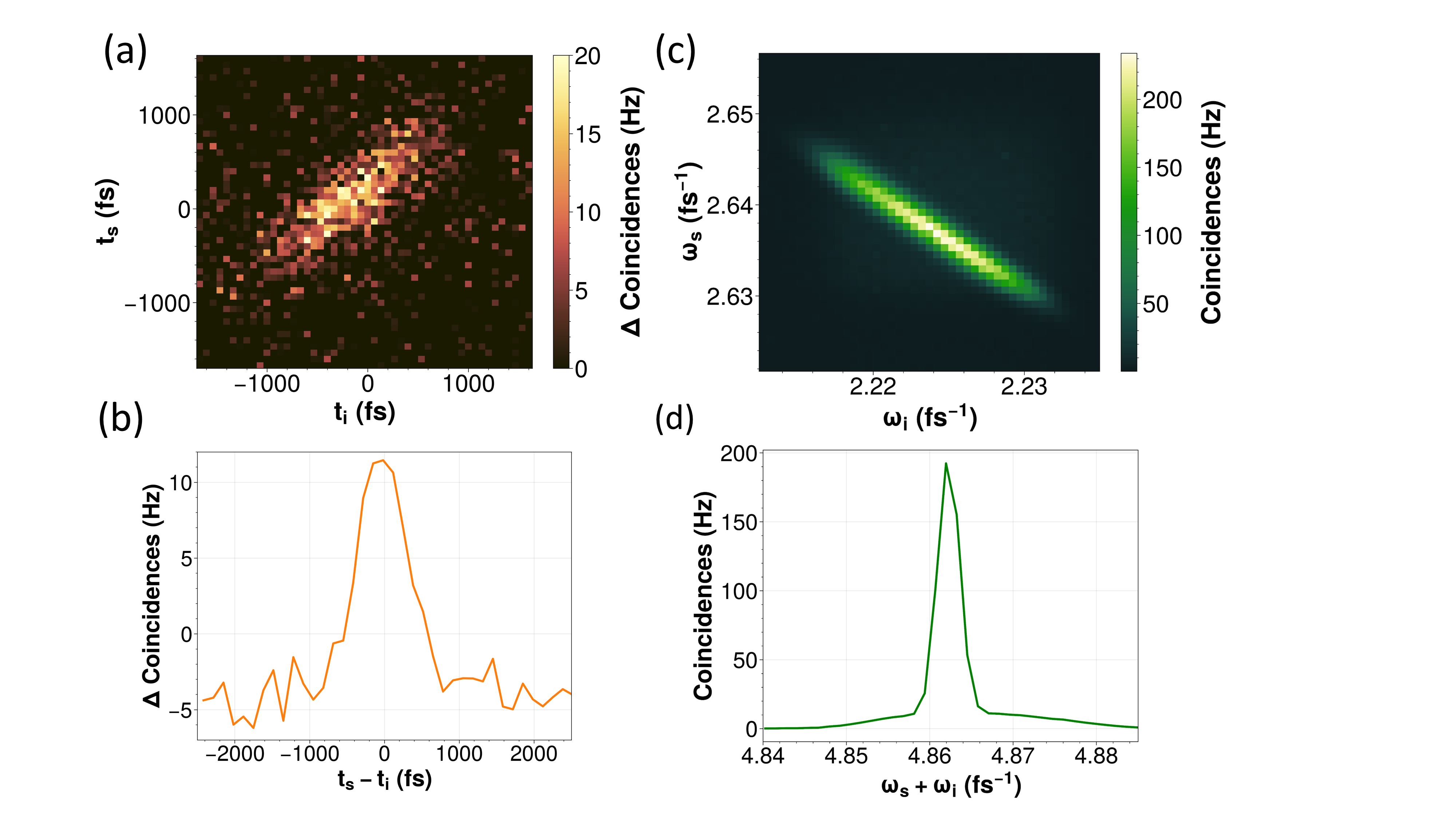}
\caption{Experimental characterization of the photon-pair temporal and spectral correlations. (a) Joint temporal intensity, (b) joint spectral intensity, (c) Cross-sectional slices of (a) about the $t_s= -t_i$ axis, and (d) Cross-sectional slices of (b) about the $\omega_s = \omega_i$ axis. For both (b) and (d), multiple slices were taken through the distributions and averaged together. The time-of-arrival of the two photons are positively correlated, while their frequencies are anticorrelated. Negative coincidence values in (c) are possible only because of the JTI background subtraction demonstrated in \figref{fig:background_subtraction}. Gaussian fitting to plots (c) and (d) yields $\Delta (t_s - t_i) = (340 \pm 30)$\,fs and $\Delta (\omega_s + \omega_i) = (0.00141 \pm 0.00002)$\,fs$^{-1}$. Together, these quantities demonstrate entanglement with $\Delta (t_s - t_i)\Delta (\omega_s + \omega_i) = (0.48 \pm 0.04)$ which is less than one by 13 standard deviations.
}
\label{fig:JTI_JSI}
\end{figure}

We define the maximum achieved gating efficiency to be the fraction of photons gated out of all possible photons, acknowledging that the pump samples a short window in time and photons have varying arrival times related to the width of the downconversion crystal and the probabilistic pair-generation. The maximum achieved gating efficiency is estimated to be approximately 16\% at the peak of the JTI. The total efficiency of the entire temporal coincidence measurement which combines coupling losses from the grating compressor, both OKS coupling losses, both final SMF coupling losses, and gating efficiency is approximately 0.01\%.

%%%%%%%%%%%%%%%%%%%%%%%%%%%%%%%%%%%%%%%%%%%%%%%%%%%%%%%%%%%%%%%%%%%%%%%%%%%%%%%%%%%%%%%%%%%%%%%%%%%%%%%%%%%%%%%%%%%%%%%%%%%%%%
%\section{Experimental Results}

%The measured JTI of the two-photon entangled state is presented in \figref{fig:background_subtraction}(c). 

We present both the raw data and noise from the OKS temporal measurement in \figref{fig:background_subtraction}. Each pixel of the images in (a) and (b) is a coincidence measurement between the signal OKS and idler OKS with relative delays $t_s$ and $t_i$. \figref{fig:background_subtraction}(a) shows the raw data which includes the two-photon correlations as well as the unwanted background consisting of accidental coincidences and constant pump leakage, while \figref{fig:background_subtraction}(b) is an estimate of background. The coincidence rates between the gated signal and idler photons are a function of electronic delay between the two detectors. Accidental coincidence peaks occur every 12.5\,ns from the relative zero delay, corresponding to the repetition rate of the laser. The peak directly following the relative zero delay is used to estimate accidental coincidences plus the constant pump leakage. The background profile in \figref{fig:background_subtraction}(b) reveals a horizontal ``stripe'' pattern in the accidental coincidences. The relative noise difference of the signal and idler OKS is apparent in \figref{fig:background_subtraction}(c). The idler noise rate is 480\,kHz counts compared to the 200\,kHz counts for the signal channel. This is because Raman scattering is higher on the low-frequency (Stokes) side of the pump pulse. 

Both temporal and spectral correlations are presented in \figref{fig:JTI_JSI}. The JTI in \figref{fig:JTI_JSI}(a) is obtained by pixel-wise subtraction of the raw data and background estimation from \figref{fig:background_subtraction}(a) and (b). Note, there are two timescales of importance: A cross-sectional slice of the JTI corresponds to the temporal width of an individual photon with 320 $\pm$ 30\,fs for the signal, and 290 $\pm$ 30\,fs for the idler, while the marginals of the JTI correspond to the uncertainty in arrival time which we measure to be 470 $\pm$ 30\,fs for signal and 520 $\pm$ 30\,fs for idler. This can be visualized in \figref{fig:walkoff} where the blue photon pulse has a much wider envelope than any given photon, shown as dotted lines, in the temporal distribution. Spectral measurements taken with single photon scanning monochromators exhibit low noise and therefore do not require the background subtraction procedure discussed for the temporal measurements. 

Measurement uncertainty takes into account the Poissonian statistics of the photon counting as well as the effect of background subtraction. The JTI is measured by subtracting the estimated background \figref{fig:background_subtraction}(b) from the raw data \figref{fig:background_subtraction}(a). Because the data and background are very similar in magnitude, the JTI measurement is highly sensitive to the estimated background. To quantify this effect, we multiply the estimated background by a variable scaling factor to determine the dependence of $\Delta (t_s - t_i)$ on the background level. This variation is included in the reported measurement uncertainty.

%Measurement uncertainty takes into account the Poissonian statistics of photon counting and the sensitivity of background subtraction. The raw data and background estimation in \figref{fig:background_subtraction}(a) and (b) are very similar in magnitude and a direct subtraction reveals the JTI we presented. To understand the sensitivity of background subtraction, the background estimation was weighted by a constant which we varied. Different subtraction weights slightly varied the measured widths and Gaussian fits and this is reflected in our reported measurement uncertainty.   

The estimate in Eq.\ref{jti_width_model} gives $\Delta (t_s - t_i) = (430 \pm 30)$\,fs. Experimental data reveals a somewhat smaller width of $\Delta (t_s - t_i) = (340 \pm 30)$\,fs. Many cross-sectional slices of the JTI about the $t_s = -t_i$ axis and JSI about the $\omega_s = \omega_i$ axis are averaged and shown in \figref{fig:JTI_JSI}(c),(d). Gaussian fits to these curves show an entanglement witness value of $\Delta (t_s - t_i) \Delta (\omega_s + \omega_i) = (340 \pm 30)$\,fs $(0.00141 \pm 0.00002)$\,fs$^{-1}$ $= 0.48 \pm 0.04$. This demonstrates a violation of Eq.~\ref{time_bandwidth_inequality} by 13 standard deviations indicating entanglement in the energy-time degree of freedom. 

%%%%%%%%%%%%%%%%%%%%%%%%%%%%%%%%%%%%%%%%%%%%%%%%%%%%%%%%%%%%%%%%%%%%%%%%%%%%%%%%%%%%%%%%%%%%%%%%%%%%%%%%%%%%%%%%%%%%%%%%%%%%%%

%\section{Conclusion}

%Future OKS pulse metrology experiments might operate with lower Raman scattering noise levels by using a pump pulse frequency lower than both the signal and idler pulses. Longer interaction lengths for improved efficiency could be used, while still maintaining sufficient temporal resolution, by use of dispersion-engineered photonic crystal fibers for pump-photon group velocity matching~\cite{sagnac_kerr_switch}. The measurement technique presented in this work can also be extended to reconstruct the complete wavefunction of two-photon energy-time entangled states. Lastly, we measured at the transmitted port of a PBS at each OKS, but by using both ports, an OKS could be used for fast rerouting of single photons where the path after the PBS is different for switched versus unswitched light. 

Our experimental parameters were based on a compromise of the various factors that influence efficiency, noise, and temporal resolution in an OKS. The OKS efficiency increases with pump intensity and fiber length; however, increasing these parameters also increases the generation rate of noise photons by self-phase modulation and spontaneous Raman scattering. In general, a large frequency difference between the pump and photon pulses will limit the generation of these noise photons at the signal and idler frequencies. In our normally dispersive SMF, an increased frequency difference increases the pump-photon group velocity walk-off and thus reduces the temporal resolution. We operated our experiment with a frequency difference $\Delta\omega=32$\,THz, fiber length $L=35$\,mm, and pump power $P=800$\,mW, yielding sufficient efficiency, signal-to-noise ratio, and temporal resolution, to measure energy-time entanglement. Future OKS pulse metrology experiments might operate with lower Raman scattering noise levels by using a pump pulse frequency lower than both the signal and idler pulses. The use of dispersion-engineered photonic crystal fibers for pump-photon group velocity matching~\cite{sagnac_kerr_switch} would enable the use of longer fibers for increased efficiency, while retaining optimal temporal resolution.

In this work, we used optical Kerr shutters to directly measure the JTI of two energy-time entangled photons. Correlations were measured with a temporal resolution of 320 $\pm$ 30\,fs and 290 $\pm$ 30\,fs for signal and idler photons, respectively, which provided a sufficient gating resolution to violate a time-bandwidth inequality and therefore witness energy-time entanglement. With the capability of distinguishing events less than one picosecond apart, and therefore outside the temporal resolution of current detector speeds, the optical Kerr shutter is a valuable addition to the available methods in ultrafast quantum optics control and metrology.

%%%%%%%%%%%%%%%%%%%%%%%%%%%%%%%%%%%%%%%%%%%%%%%%%%%%%%%%%%%%%%%%%%%%%%%%%%%%%%%%%%%%%%%%%%%%%%%%%%%%%%%%%%%%%%%%%%%%%%%%%%%%%%

%\section*{Acknowledgments}
The authors would like to thank John Donohue and Jean-Philippe MacLean for helpful discussions and assistance. This research was supported in part by the Natural Sciences and Engineering Research Council of Canada (NSERC), Canada Research Chairs, Industry Canada, the Canada Foundation for Innovation (CFI), Ontario Research Fund (ORF), and the Canada First Research Excellence Fund (CFREF).

%%%%%%%%%%%%%%%%%%%%%%%%%%%%%%%%%%%%%%%%%%%%%%%%%%%%%%%%%%%%%%%%%%%%%%%%%%%%%%%%%%%%%%%%%%%%%%%%%%%%%%%%%%%%%%%%%%%%%%%%%%%%%%

\bibliography{kerr_switch}

\end{document}